\documentclass{article}

\usepackage{arxiv}

\usepackage[utf8]{inputenc} 
\usepackage[T1]{fontenc}    
\usepackage{hyperref}       
\usepackage{url}            
\usepackage{booktabs}       
\usepackage{amsfonts}       
\usepackage{nicefrac}       
\usepackage{microtype}      
\usepackage{lipsum}		
\usepackage{graphicx}
\usepackage{authblk}
\usepackage{float}
\usepackage{xcolor}

\usepackage[backend=bibtex,style=numeric]{biblatex} 
\title{Integrating Large Language Models for Genetic Variant Classification}

\author[1]{Youssef Boulaimen}
\author[1]{Gabriele Fossi}
\author[1]{Leila Outemzabet}
\author[1]{Nathalie Jeanray}
\author[1]{Oleksandr Levenets}
\author[1]{Stéphane Gerart}
\author[1]{Sébastien Vachenc}
\author[1,2]{Salvatore Raieli}
\author[1]{Joanna Giemza}

\affil[1]{Oncodesign Precision Medicine, 18 rue Jean Mazen, 21000 Dijon, France}
\affil[2]{Corresponding author: \texttt{sraieli@oncodesign.com}}

\renewcommand{\shorttitle}{LLM for variant classifications}

\hypersetup{
pdftitle={Integrating Large Language Models for Genetic Variant Classification},
pdfsubject={q-bio.NC, q-bio.QM},
pdfauthor={Boulaimen, Fossi, Outzemabet, Jeanrey, Levenets, Vachenc, Gerart, Raieli, Giemza},
pdfkeywords={AI, LLM, LLM in biology, Variable uncertain significance},
}

\begin{document}
\maketitle

\begin{abstract}
	\
    The classification of genetic variants, particularly Variants of Uncertain Significance (VUS), poses a significant challenge in clinical genetics and precision medicine. Large Language Models (LLMs) have emerged as transformative tools in this realm. These models can uncover intricate patterns and predictive insights that traditional methods might miss, thus enhancing the predictive accuracy of genetic variant pathogenicity.
    This study investigates the integration of state-of-the-art LLMs, including GPN-MSA, ESM1b, and AlphaMissense, which leverage DNA and protein sequence data alongside structural insights to form a comprehensive analytical framework for variant classification. Our approach evaluates these integrated models using the well-annotated ProteinGym and ClinVar datasets, setting new benchmarks in classification performance. The models were rigorously tested on a set of challenging variants, demonstrating substantial improvements over existing state-of-the-art tools, especially in handling ambiguous and clinically uncertain variants.
    The results of this research underline the efficacy of combining multiple modeling approaches to significantly refine the accuracy and reliability of genetic variant classification systems. These findings support the deployment of these advanced computational models in clinical environments, where they can significantly enhance the diagnostic processes for genetic disorders, ultimately pushing the boundaries of personalized medicine by offering more detailed and actionable genetic insights.

\end{abstract}


\section{Introduction}
\
The emergence of Next Generation Sequencing (NGS) (reviewed in \cite{qin_next-generation_2019}) has transformed the realm of genomics, enabling the sequencing of millions of DNA fragments. However, the interpretation of the NGS results poses significant challenges as the vast majority of identified variants are of unknown significance (VUS)\cite{fayer_closing_2021,chennen_mistic_2020,cheng_accurate_2023}. An accurate prediction of such variants can pave the way to a better understanding of disease mechanisms, enabling personalized medicine and the discovery of new therapeutic targets.

Over the years, many computational tools and datasets have been developed to help predict the effects of variants. Early tools like PolyPhen and SIFT used sequence homology and protein structure information to predict the impact of missense mutations \cite{adzhubei_predicting_2013,ng_sift_2003}. Other models, such as CADD \cite{kircher_general_2014}, combine multiple annotations into a single score to indicate variant pathogenicity.

The promising results of Large Language Models (LLMs) in Natural Language Processing (NLP) tasks have led to their adaptations in the fields of genomics and proteomics. LLMs are complex models that use the Transformer architecture\cite{vaswani_attention_2023}. One particular component of the Transformer architecture is self-attention, which enables the model to weigh the importance of different parts of the input data dynamically. This mechanism allows the models to consider the entire sequence context, making it particularly effective in handling long-range dependencies and interactions within the data. A remarkable example of the high potential of LLMs in proteomics is ESMFold, a protein language model that can predict protein structures using protein sequences\cite{lin_language_2022}. 

In variant effect prediction (VEP), exploiting the capabilities of self-attention can be beneficial as it allows the model to account for not only specific mutations but also the entire genetic background and associated protein sequences, providing a comprehensive view of the molecular context. 
LLMs such as GPN-MSA, ESM1b, and Alphamissense have shown promise in predicting variant pathogenicity.
GPN-MSA is a DNA language model trained on MSA (Multiple Sequence Alignment) of 100 species which leverages evolutionary information in predicting pathogenicity scores for all possible nucleotide substitutions in the genome\cite{noauthor_gpn-msa_nodate}. ESM1b is a protein language model that predicts the pathogenicity for all 20 possible amino acids, without relying on homology and taking into account all protein isoforms\cite{brandes_genome-wide_2023}. As for Alphamissense, it is first trained to predict protein structures from the protein sequences and later fine-tuned for pathogenicity prediction\cite{cheng_accurate_2023}. These models are considered as state of the art in VEP.

We hypothesize that integrating these models may offer significant advantages. By combining their predictions, we can not only capitalize on their strengths and address their weaknesses but also provide a more comprehensive prediction leveraging both DNA and protein data. We adopted this integrative approach, using machine learning models for the combination of the scores, in order to develop a more accurate and comprehensive tool for predicting variant pathogenicity.


\section{Materials \& Methods}
\label{sec:material and methods}
\subsection{Data}
\
In this study, we utilized the ProteinGym dataset\cite{noauthor_proteingym_nodate}, a comprehensive resource developed to facilitate the evaluation of mutation effect predictors. This dataset is divided into two primary benchmarks: substitution benchmark and indel benchmark. For our analysis, we focused exclusively on the substitution benchmark of the \href{https://proteingym.org/download}{ProteinGym dataset} (accessed on 3/22/24), which includes approximately 2.7 million missense variants characterized across 217 Deep Mutational Scanning (DMS) assays\cite{fowler_deep_2014} and encompasses data on 2,525 clinical variants. We examined specifically two distinct segments within this benchmark: the clinical variants substitutions dataset and the raw substitutions dataset. The raw substitution dataset is extensive and contains 61 columns. Crucial to our study are the columns indicating the chromosome and the exact genomic location of each variant, as well as reference and alternative alleles, which detail the nucleotide changes. Also integral to our analysis are the columns detailing the corresponding protein position and the amino acid changes. These are linked via the transcript ID, which connects the genomic data to specific protein transcripts, thereby facilitating cross-references between the genetic and protein data. Moreover, the dataset includes columns that categorize the clinical significance of each variant, classifying them as either pathogenic or benign.
The clinical substitutions dataset contains the transcript ID to ensure consistent referencing across the datasets. It records both the position within the protein and the reference and alternate amino acids involved in each substitution. Additionally, it provides the sequences before and after mutations, along with the DMS\_bin\_score that classifies each protein substitution as benign or pathogenic.

GPN-MSA’s HuggingFace \href{https://huggingface.co/datasets/songlab/gpn-msa-hg38-scores/tree/main}{repository} (accessed on 3/12/24) provides predictions for all possible SNPs in the human genome. Using the chromosomes and positions of the variants from the raw substitutions dataset, we queried the scores for all three possible nucleotide substitutions using Tabix\cite{li_tabix_2011}. The lower the score of GPN, the more pathogenic the variant.

We employed the ProteinGym substitution dataset to compute the ESM1b scores, which include reference protein sequences along with detailed mutation information such as positions and the specific amino acids involved. The ESM1b code was sourced from its \href{https://github.com/ntranoslab/esm-variants}{GitHub repository}. This model takes protein sequences as input and produces scores for all 20 possible amino acid substitutions at each position within the sequence, which can lead to extensive output files and considerable processing times. To streamline this process, we modified the ESM1b code to focus on scoring only the specified mutation positions from the dataset. This targeted approach significantly reduced both the output file size and the processing time. The lower the score of ESM1b, the more pathogenic the variant. A Log Likelihood Ratio (LLR) threshold of -7.5 was used to distinguish between pathogenic and benign variants\cite{brandes_genome-wide_2023}.

For our analysis, we also utilized predictions from AlphaMissense, accessible through the file \href{https://zenodo.org/records/8360242}{AlphaMissense-aa-substitutions.tsv.gz} (4/8/2024). This dataset contains predictions for all conceivable single amino acid substitutions within 20,000 UniProt canonical isoforms, totaling approximately 216 million protein variants.
Integrating AlphaMissense predictions posed significant challenges due to discrepancies in protein identifier systems. AlphaMissense uses UniProt accession numbers, whereas the ProteinGym dataset relies on NCBI’s RefSeq protein IDs. To address this, we utilized the UniProt ID mapping tool to align the datasets, successfully mapping 2,415 out of the 2,525 proteins from ProteinGym. This mapping allowed us to accurately link UniProt accession numbers to the corresponding mutations in ProteinGym and retrieve the necessary AlphaMissense scores for our analysis.


\subsection{Data Processing}
\

\begin{figure} [!h]
    \centering
    \includegraphics[width=1\linewidth]{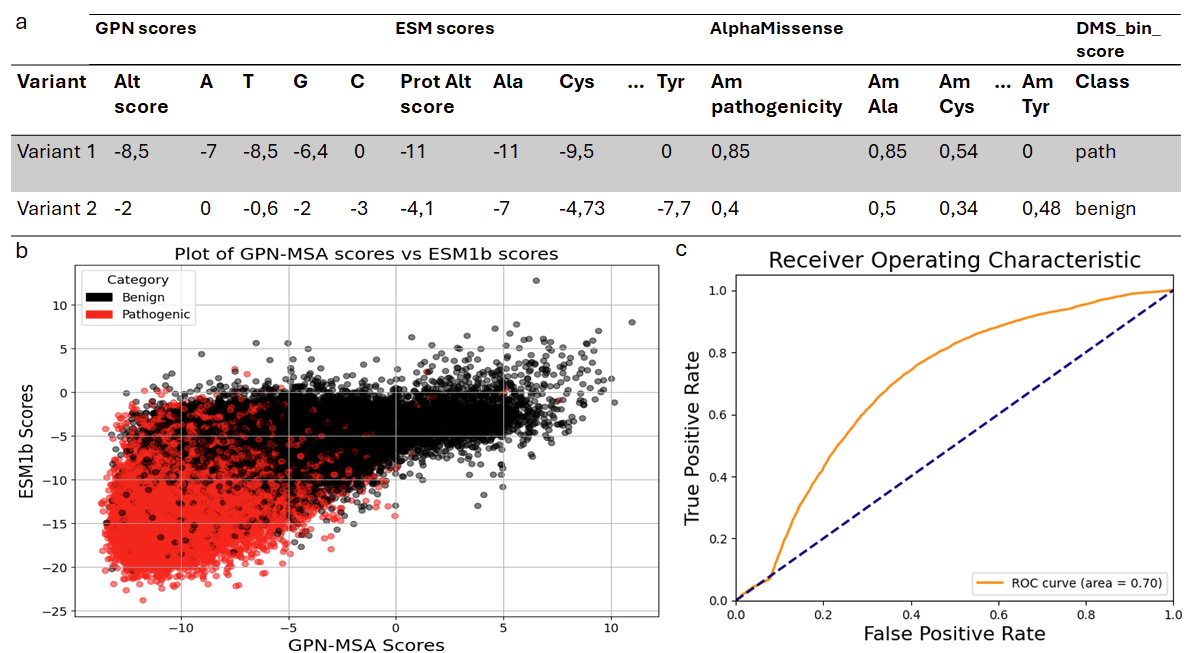}
    \caption{Dataset presentation: 
    \textbf{a. Dataset Samples:} This table provides a representative sample of the dataset utilized for this study, showcasing both observed and potential mutation scores derived from three distinct models: GPN scores, ESM scores, and AlphaMissense scores. Each entry represents the score assigned by the respective model to various attributes such as nucleotide changes (A, T, G, C) and amino acid substitutions (e.g., Ala, Cys, Tyr), as well as the observed scores in columns Alt\_score, Prot\_Alt\_score, and Am\_pathogenicity. Nucleotides and Amino Acids with a score of value 0 correspond to the reference allele or protein. The "DMS\_bin\_score" column indicates the clinical classification of the mutation as either "pathogenic" or "benign." 
    \textbf{b. Dataset Visualization:} This plot represents the distribution of the variants’ observed scores of GPN-MSA as Alt Score on the X-axis and ESM1b as Prot Alt Score on the Y-axis. Red data points represent the variants classified as pathogenic by DMS\_bin\_score whereas the black data points are classified as benign. 
    \textbf{c. Optimal Threshold for GPN:} This ROC curve illustrates the performance of the GPN model in discriminating between pathogenic and benign genetic variants. The x-axis represents the False Positive Rate (FPR), and the y-axis represents the True Positive Rate (TPR), across various threshold levels. The orange line depicts the actual ROC curve, which shows how the TPR and FPR change with different thresholds. The area under the curve (AUC) is 0.70, indicating the model's overall ability to distinguish between the classes; a value of 1.0 represents a perfect classifier, and a value of 0.5 represents a random guess. The dashed blue line represents the line of no discrimination, which serves as a baseline comparison. The optimal threshold for classification is found by maximizing the difference between TPR and FPR.
    }
    \label{fig1}
\end{figure}
This analysis differentiated between two types of scores. Firstly, the observed mutation scores, are calculated for mutations that are actually present in the dataset and represent clinically observed mutations in genomic or protein sequences. These scores provide direct insights into the impact of a specific, known mutation for one alternative nucleotide or amino acid (Fig.1.a columns Alt score, Prot Alt score, and AM pathogenicity). These observed mutations have an experimental annotation in the DMS\_bin\_score column. Secondly, the potential mutation scores, which speculate on the theoretical impact of all conceivable mutations at each position within the genome or protein sequence. For GPN-MSA, this involves generating four potential scores corresponding to the four possible nucleotide changes at each genomic position. In the case of ESM1b and AlphaMissense, scores are generated for each of the 20 possible amino acid substitutions at each position in the protein sequence ((Fig.1.a columns A, T, Ala, Am Ala...). In cases where the reference and alternative alleles or amino acids are identical, a score of zero is assigned, reflecting no change or impact due to the mutation. The potential mutations do not have a pathogenicity classification, except for the one corresponding to the observed mutation.

The first step of the data processing was merging the GPN-MSA and ESM1b scores using the transcript ID and protein information from both ProteinGym datasets, thus obtaining a dataset of 59,593 rows. This dataset was used for the preparation of the training and testing sets for the deep learning models. The splitting was performed in a manner that keeps the most ambiguous and difficult-to-predict data points in the test set by selecting a threshold for both scores. After visualizing the distribution (Fig.1), we selected the variants with scores between -5 to -10.

AlphaMissense scores were added later, resulting in a small reduction of the dataset to 49,554 rows due to the proteins lost during the mapping. A new test set was generated by merging the previous test set with the new dataset containing the 3 scores. Using the threshold resulted in a test set of 16165 rows and a training set of 33,389 rows with balanced proportions of 16,588 pathogenic and 16,801 benign variants.

The last step was to assign a threshold for the GPN-MSA scores to enable the comparison between all models. To find the optimal threshold, one approach is to maximize the difference between True Positive Rate (TPR) and False Positive Rate (FPR). This is typically done by calculating TPR and FPR for each possible threshold using the Receiver Operating Characteristic (ROC) curve, and then identifying the point where the difference between TPR and FPR is the greatest. This index corresponds to the optimal threshold from the evaluated thresholds array. For the GPN-MSA scores, this method was used to identify a threshold that optimally differentiates between pathogenic and benign variants, ensuring accurate and clinically relevant comparisons across all models. The threshold found for the GPN-MSA scores was -7 with an optimal FPR of 0.41 and optimal TPR of 0.759.
For the other models, AlphaMissense predictions were taken directly from its am\_class output, which labels variants as either Pathogenic, Benign or Ambiguous. For ESM1b, variants with a score of -7.5 or below were considered pathogenic as described in the paper\cite{brandes_genome-wide_2023}.

\subsection{Model Architectures}
Various machine learning models were used for the training, namely XGBoost (XGB)\cite{chen_xgboost_2016}, Random Forest (RF)\cite{breiman_random_2001}, and Neural Networks. All models were trained using different sets of pathogenicity scores as features and the DMS\_bin\_score as the target variable.
The ensemble models such as XGB and RF were trained using default parameters, as the fine-tuning of such models requires using grid search, which demands extensive amounts of time for limited improvements.
As for Neural Networks, several architectures were employed. As the protein and DNA models provide a different number of scores, we decided to explore both Multi-input Neural Networks that take each score separately and Single-input Neural Networks that take all of the scores altogether. The architectures and parameters were explored and the optimal values were selected through a process of trial and error. The architectures and parameters were systematically optimized through an iterative process of trial and error to determine the optimal configuration.

\subsubsection{Single input Neural Networks}
The model architecture included a single input layer to handle the scores.
 This input was passed through a Dense layer with 64 units and a LeakyReLU activation function. A Dropout layer with a rate of 0.5 followed to prevent overfitting. The dropout layer was connected to another Dense layer with 128 units and a LeakyReLU activation function. The final output layer was a single unit Dense layer with a sigmoid activation function for binary classification.

The model was compiled using the Stochastic Gradient Descent (SGD) optimizer with a learning rate of 0.001 and binary cross-entropy loss function. Training was performed for up to 350 epochs with a batch size of 32. Early Stopping with a patience of 10 epochs and ReduceLROnPlateau with a factor of 0.2 and threshold of 0.0001 were employed to prevent overfitting and adjust the learning rate, respectively\cite{abadi_tensorflow_2016}.

\subsubsection{Multi-input Neural Networks}

The model included three input layers to handle the different scores. Each branch began with a Dense layer of 64 units, followed by Batch Normalization and ReLU activation.
The outputs from these three branches were concatenated into a single tensor. This concatenated tensor was then passed through two Dense layers with 256 and 128 units, respectively. Each dense layer was followed by Batch Normalization, ReLU activation, and Dropout with a rate of 0.5 to prevent overfitting. The final output layer was a single unit Dense layer with a sigmoid activation function, appropriate for binary classification tasks.
The model was compiled using the Stochastic Gradient Descent (SGD) optimizer with a learning rate of 0.01 and the binary cross-entropy loss function. Training was conducted for up to 350 epochs with a batch size of 32. To prevent overfitting and adjust the learning rate during training, Early Stopping with a patience of 50 epochs and ReduceLROnPlateau with a factor of 0.2 and threshold of 0.0001 were utilized.

\subsection{Case study methodology}
Case studies were conducted on select variants to confirm the model's prediction and showcase its efficiency as described in 3.4.
We followed a specific methodology in order to extract the necessary data about these variants.
First, we look into the protein information in Uniprot using the UniProt ID in our dataset. We first note the general data about the protein such as the function and the structure. Next, we download the PDB file for the AlphaFold structure of the protein from AlphaFold's website. The PDB file is then used to visualize the protein structure with Pymol. Pymol enables to change the specific residue in a specific position to another residue and observe the changes in the protein structure using the Mutagenesis panel. We use this to obtain the structure for the protein with the mutation we need. This way it is possible to visualize both the wild-type protein's structure as well as the mutated protein. Later we explore the interactions of the WT and mutated residues with the neighbouring structures in a radius of 3.5 Angstroms. This enables us to hypothesize on the possible outcomes of the mutation.
Next, we investigate the Disease \& Variants section which contains the known diseases the protein is involved in and the variants that are possibly implicated. For the well-annotated and well-studied variants, links for scientific papers are provided which are also investigated.
In case the mutation we investigate is not included in the Disease \& Variants section, the variant viewer panel in UniProt. This panel contains information on several mutations in the protein such as the effect of the mutation from different databases but also potential diseases in which the mutation can be involved. The main databases used are usually ClinVar, gnomAD, and dbSNP.
Finally, we explore the Family \& Domains section to check whether the mutation is involved in a functional domain of the protein. This section also provides scientific papers of studies conducted on specific regions of the proteins.

\subsection{Technical Details}

The computational analysis was conducted using Python 3.8 in a Jupyter Notebook environment. The platform used was a Linux operating system (Linux-5.15.0-1044-nvidia-x86\_64-with-glibc2.27). The hardware specifications include a 64-bit processor architecture (x86\_64) with 40 physical cores, providing substantial computational power for parallel processing tasks. The system was equipped with 503 GB of total memory (RAM). The total disk space available was 438 GB.
A key feature of the setup is the presence of eight NVIDIA Tesla V100-SXM2-32GB GPUs, each with 32 GB of dedicated memory. The CUDA version employed was 12.3, supported by NVIDIA driver version 545.23.08.

\section{Results}
The evaluation of several machine learning models was conducted to assess their effectiveness in predicting the pathogenicity of genetic variants. In this section, we will describe a correlation analysis between the scores of GPN-MSA, ESM1b, and AlphaMissense, benchmarking of the various models and a selection of the optimal set of features. Next, we will compare our best model’s performance to state-of-the-art tools to contextualize its performance. Finally, we performed case studies to further showcase our model's utility in real-world applications

We utilized a dataset composed of various genetic scores derived from GPN-MSA, ESM1b, and AlphaMissense. These scores were the primary inputs for the predictive models.
Four different feature sets were considered as inputs for the predictive models:
\begin{itemize}
    \item Potential scores from GPN-MSA and ESM1b: This feature set contains all four possible predictions from GPN-MSA and twenty predictions from ESM1b. It will test the impact of integrating both DNA and protein data and will also serve as a basis for comparison with other models that will incorporate Alphamissense scores.
    \item Potential scores from GPN-MSA, ESM1b, and AlphaMissense: This feature set contains all four possible predictions from GPN-MSA and twenty predictions from ESM1b and Alphamissense. This feature set will assess the utility of adding Alphamissense scores to the model.
    \item Observed Scores from GPN-MSA, ESM1b, and AlphaMissense: This feature set contains only one score for each model. This score corresponds to the score of the observed mutation in the dataset. Using this feature set we will test the necessity of using all the potential scores.
    \item Observed and Potential Scores from GPN-MSA, ESM1b, and AlphaMissense: This feature set includes the scores for observed mutations for each model alongside all the possible model predictions. Observed scores represent the actual clinical mutations identified in the dataset, reflecting real-world genetic variations with known clinical significance. Potential scores, on the other hand, encompass a broad range of hypothetical mutations that provide a comprehensive view of possible genetic variations. Including both sets of scores in the feature set introduces a duplication of the observed scores, effectively weighting them more heavily in the model. This feature engineering approach is designed to emphasize the real, clinically validated mutations captured by the observed scores, thereby potentially improving the model's performance.
\end{itemize}
The models were trained on a subset of the data and evaluated on a separate test set specifically chosen to include the most ambiguous and challenging variants. This approach was used to evaluate the model performance as strictly as possible, ensuring that the models are robust and effective even under the most difficult conditions. The test set composition included a balanced mix of 7902 pathogenic and 8263 benign variants.

\subsection{Correlation Analysis}

The correlation analysis quantified the degree of linear association between the observed scores from GPN-MSA, ESM1b, and AlphaMissense using Pearson correlation coefficients. This analysis provides insights into the prediction trends among these models, evaluating the necessity for an integrated model approach.
GPN-MSA and ESM1b show a positive correlation (0.6779), suggesting these models tend to align in their predictions. In contrast, GPN-MSA and AlphaMissense have a negative correlation (-0.7259), and ESM1b and AlphaMissense exhibit an even stronger negative correlation (-0.8104).
The negative correlations between AlphaMissense and the other two models (GPN-MSA and ESM1b) are expected, given that AlphaMissense assigns higher scores to pathogenic variants while the other two models assign higher scores to benign variants. The stronger correlation between ESM1b and AlphaMissense is due to their focus on protein-level pathogenicity predictions, while GPN-MSA predicts at the DNA level, explaining its lower correlation with both models.
Overall, the results justify the need for an integrated model, as the individual models capture different aspects of the data, providing a more comprehensive and nuanced analysis when combined.
\begin{figure}[!h]

    \centering
    \includegraphics[width=0.75\linewidth]{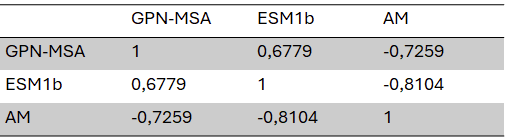}
    \caption{\textbf{Correlation Analysis} 
    \textbf{} This table illustrates the correlation matrix between the observed scores from the GPN-MSA, ESM1b, and AlphaMissense models. The correlation coefficients quantify the degree to which these models agree or disagree on the pathogenic potential of the mutations, providing insight into their comparative analytical behaviors.} 
\end{figure}

\subsection{Model performances}

\begin{figure}[!h]
     \centering
     \includegraphics[width=1\linewidth]{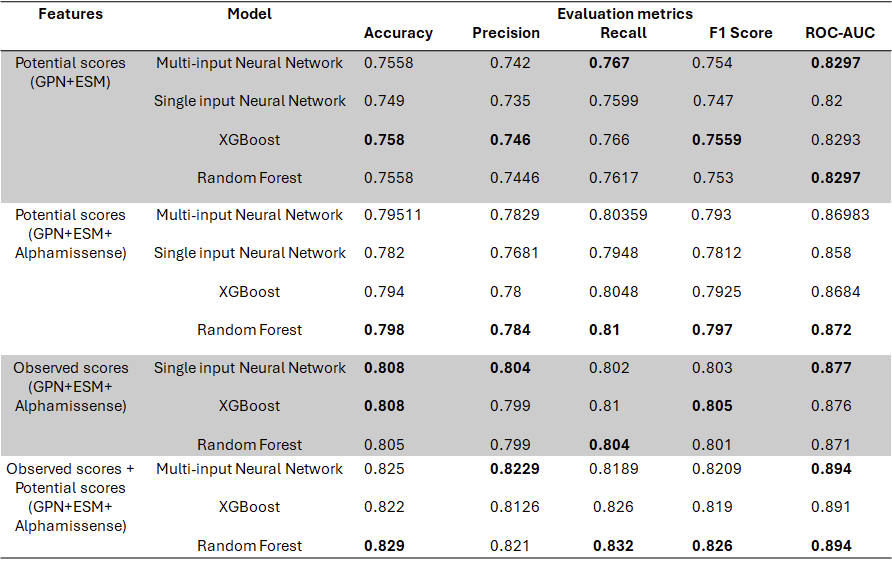}
     \caption{\textbf{Comparative performance of Machine Learning models in genetic variant classification:}
     \ This table presents the benchmarking results of different machine learning models using various combinations of features derived from GPN-MSA, ESM1b, and AlphaMissense. The models evaluated include multi-input neural networks, single-input neural networks, XGBoost, and Random Forest, each tested across four distinct feature sets: GPN+ESM potential scores, GPN+ESM+AlphaMissense potential scores, observed scores from GPN+ESM+AlphaMissense, and a combination of observed and potential scores from GPN+ESM+AlphaMissense.
}
     \label{fig:enter-label}
 \end{figure}
 
Here we evaluate the performances of the models across the different feature sets. The model performances are detailed in Fig.3. Overall, the choice of the models didn’t have a significant impact on the results. However, the evaluation revealed varying performance across the different feature configurations.
All models showed similar performance with the GPN+ESM feature set, with accuracies around 0.75 and ROC-AUC just below 0.83. The multi-input neural network and random forest slightly outperformed the XGBoost and single-input neural network in terms of ROC-AUC.
The incorporation of AlphaMissense scores improved performance across all models, particularly the Random Forest model with a ROC-AUC of 0.872.
The single-input neural network excelled with the Observed Scores configuration, achieving the highest accuracy 0.808 and ROC-AUC of 0.877.
 The feature set that included both Observed Scores and Potential Scores provided the best overall results, particularly for the multi-input neural network and random forest, both achieving high accuracies and ROC-AUC scores above 0.89.

\vspace{4.5cm}

\subsection{Performance comparison state of the art models
}
Next, we will compare the Multi-input Neural Network model trained on Observed + Potential scores to state-of-the-art tools. This analysis aims to assess the real-world applicability of our approach, particularly in its ability to accurately predict Variants of Uncertain Significance (VUS). By benchmarking our model against existing leading tools, we aim to demonstrate its effectiveness and potential advantages in clinical and research settings. This comparison will help us understand how well our integrated model performs in practical scenarios, ensuring its utility in improving genetic variant classification.
\subsubsection{Performance comparison state of the art models: ProteinGym
}

In this study, the Multi-input NN model trained on Observed+Potential scores was evaluated alongside AlphaMissense, GPN-MSA, and ESM1b, using the test dataset of 16,165 genetic variants. This dataset was specifically chosen to include the most ambiguous and challenging variants to evaluate model performance as strictly as possible, ensuring robustness and effectiveness under difficult conditions. The models' performance was assessed by comparing their predictions against the experimental annotations provided in the ProteinGym's DMS\_bin\_score. For the assessment, AlphaMissense predictions were taken directly from its am\_class output, which labels variants as either Pathogenic, Benign or Ambiguous. For ESM1b, variants with a score of -7.5 or below were considered pathogenic. The threshold for GPN-MSA was set at -7, as detailed in the Materials and Methods.
\begin{figure}[!h]
    \centering
    \includegraphics[width=1\linewidth]{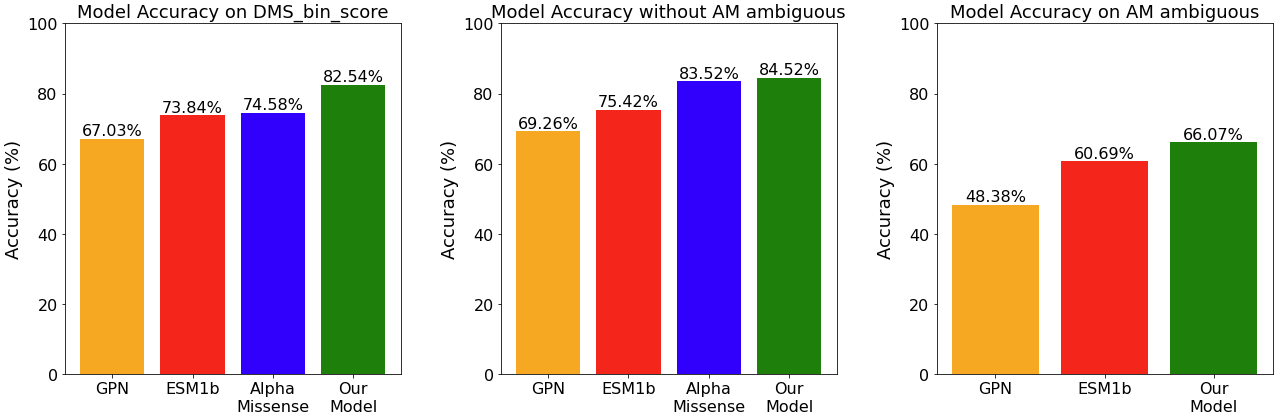}
    \caption{\textbf{Model Performance Across Different Conditions and Datasets:}
    \textbf{(a):} This panel displays the accuracy of individual models on the DMS\_bin\_score. The bars represent the accuracy of the AlphaMissense model, the integrated Model (combining GPN-MSA, ESM1b, and AlphaMissense), ESM1b, and GPN. Here, the integrated Model shows a strong performance with an accuracy of 82.54\%, followed by AlphaMissense at 74.58\%. ESM1b and GPN exhibit lower accuracies at 73.84\% and 67.03\% respectively.
    \textbf{(b):} This panel illustrates the model accuracy after removing variants classified as ambiguous by AlphaMissense. This graph provides insights into how the clarity of variant classification affects model performance. The integrated Model achieves the highest accuracy at 84.51\%, followed by AlphaMissense at 83.51\%, and ESM1b at 75.41\%. GPN shows significantly lower accuracy at 69.26\%, suggesting it is more affected by the removal of ambiguous variants compared to the other models.
    \textbf{(c):} This panel focuses specifically on the accuracy of models in predicting AlphaMissense classified ambiguous variants, highlighting the challenges in handling ambiguous genomic data. The integrated Model maintains the highest accuracy at 66.07\%, demonstrating its robustness even in uncertain conditions. ESM1b and GPN show reduced accuracies at 60.69\% and 48.38\% respectively.
    }
    \label{fig3}
\end{figure}
The initial comparison of the models was performed on all 16,165 variants of the test set (Fig.4a). 
This comparison aims to provide a baseline comparison of the models on the challenging variants of the test set. The analysis demonstrated that the integrated model outperformed the others, achieving an accuracy of 82.54\%. AlphaMissense and ESM1b showed comparable performances with accuracy of 74.58\% and 73.84\% respectively, while GPN-MSA lagged at 67.03\%.
For the second analysis, variants classified as ambiguous by AlphaMissense were excluded from the analysis, reducing the test set to 14,435 variants (Fig.4b). This comparison aimed to provide a fairer assessment by considering only two predictions, removing the variants that are difficult for Alphamissense to predict. All models displayed improved accuracy.The integrated model still outperformed AlphaMissense, achieving accuracies of 84.52\% and 83.52\% respectively.
The last analysis focused on the 1,730 variants categorized as ambiguous by AlphaMissense (Fig.4c). Here we will measure how the models perform on a smaller subset of hard-to-predict variants according to AlphaMissense. In this challenging subset, the integrated model again showed robust performance, achieving the highest accuracy of 66.07\%. ESM1b and GPN, presented accuracies of 60.69\% and 48.38\% respectively.
It's important to note that we solely relied on accuracy for model evaluation in this study. AlphaMissense outputs three classifications (Pathogenic, Benign, Ambiguous), rendering the calculation of other common metrics like ROC-AUC more complex.

\begin{figure}[!h]
    \centering
    \includegraphics[width=1\linewidth]{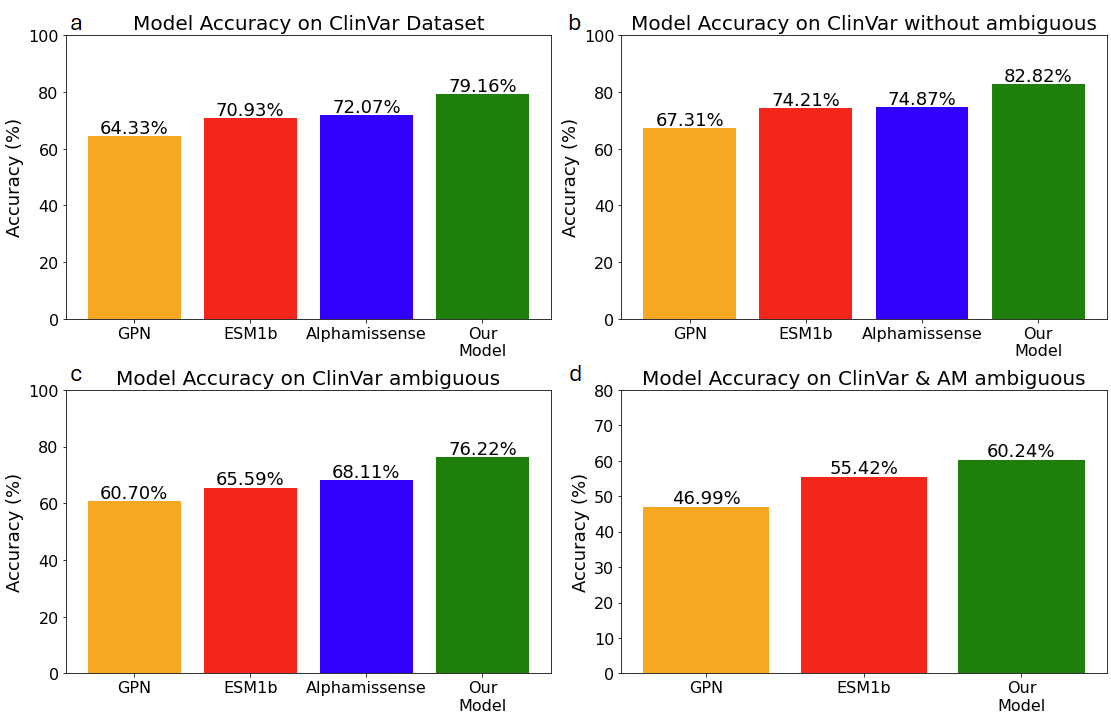}
    \caption{\textbf{Model Performance on ClinVar Dataset:}
    \textbf{(a):} This panel displays the accuracy of AlphaMissense, the integrated Model, ESM1b, and GPN when tested against the ClinVar dataset. The ClinVar dataset provides three different classes: Pathogenic, Benign, and Ambiguous. The integrated Model shows the highest accuracy at 79.16\%, followed by AlphaMissense at 72.07\%, ESM1b at 70.93\%, and GPN showing the lowest accuracy at 64.33\%.
    \textbf{(b):} This panel illustrates the accuracy of the same models on the ClinVar dataset after the removal of the 715 ambiguous variants. The performance of all models is generally improved. The integrated Model leads in accuracy at 82.82\%, demonstrating its effectiveness in classifying clearly defined genetic variants. This is followed by AlphaMissense at 74.87\%, ESM1b at 74.21\%, and GPN at 67.31\%.
    \textbf{(c):} This panel focuses on variants classified as ambiguous in the ClinVar dataset while using DMS as ground truth. The graph illustrates that the integrated Model maintains superior performance even in this subset, achieving an accuracy of 76.22\%. AlphaMissense follows at 68.11\%, ESM1b at 65.59\%, and GPN shows the least accuracy at 60.70\%.
    \textbf{(d):} This panel examines the accuracy of the models on a combined subset of variants classified as ambiguous by both ClinVar and AlphaMissense, with DMS used as ground truth for performance evaluation. The integrated Model continues to show superior performance in this challenging scenario with an accuracy of 60.24\%. This is followed by ESM1b at 55.42\%, and GPN at 46.99\%, further validating the robustness of the integrated Model.}
    \label{fig4}
\end{figure}
\subsubsection{Performance comparison state of the art models: ClinVar
}

In addition to the DMS annotation, the ClinVar\cite{landrum_clinvar_2016} classifications were added to the test dataset for further comparison. This merge implied the loss of a single row bringing the dataset to 16,164 variants.
The ClinVar data provides an array of different classifications. Variants with labels ‘Uncertain significance’ or ‘Conflicting classifications of pathogenicity’ were classified as Ambiguous. The other variants were either labeled Benign or Pathogenic. Comparing the models’ performances on the ClinVar dataset provides an additional layer of analysis on a clinically relevant dataset, offering valuable insights into the models’ effectiveness in handling Variants of Uncertain Significance (VUS).

First, we performed the comparison of each model against the ClinVar classifications Fig.5a. This comparison was motivated by the need to evaluate the overall accuracy of each model. In this analysis, AlphaMissense should have an advantage as it can directly classify ambiguous variants, whereas other models, which only provide benign or pathogenic predictions, would generate false predictions for these ambiguous variants. Despite this, the integrated model still outperformed AlphaMissense in this task. The integrated model achieved an accuracy of 79.16\%, AlphaMissense 72.07\%, ESM1b 70.93\%, and GPN 64.33\%.

Next, we tested the models’ performances on clearly classified variants by removing the 715 ambiguous variants of the ClinVar dataset Fig.5b. This analysis was conducted to provide a more straightforward comparison, focusing solely on benign and pathogenic classifications without the complexity introduced by ambiguous variants. This analysis shows an overall better accuracy across all the models compared to Fig.5.a with the integrated model outperforming the other models with an accuracy of 82.82\%, followed by AlphaMissense at 74.87\%, ESM1b at 74.21\%, and GPN at 67.31\%.

Subsequently, the performance of the models on the 715 ambiguous variants of ClinVar was assessed using the DMS\_bin\_score as the ground truth (Fig.5.c). This task aimed to evaluate how effectively each model discriminates between variants with uncertain significance. The results indicated weaker performances compared to the overall dataset (Fig.5a). Nonetheless, the integrated model again outperformed the other models with an accuracy of 76.22\%, showcasing its efficacy in classifying ambiguous variants. AlphaMissense, ESM1b, and GPN showed accuracies of 68.11\%, 65.59\%, and 60.70\%, respectively.

Finally, the performance of the integrated model, ESM1b, and GPN-MSA was assessed on a small dataset of 83 variants classified as ambiguous by both AM and ClinVar (Fig.5d). This comparison was motivated by the need to evaluate the models on the most challenging subset, where both AlphaMissense and ClinVar classifications had flagged the variants as ambiguous. The DMS\_bin\_score was used as the ground truth for this evaluation. The integrated model continued to display robust results with 60.24\% correct predictions, followed by ESM1b with 55.42\% and GPN with 43.37\%. Assessing the models on this difficult subset underscores the integrated model's capability to manage the complexities associated with ambiguous genetic data, further validating its robustness and reliability in real-world applications where uncertain classifications are prevalent.

\subsection{Case studies}

Here we investigate examples of variants classified as VUS by ClinVar and for which our model aligns with DMS\_bin\_score as opposed to the other three models. These case studies help showcase the utility of our model in real-life scenarios but also prove our model effectively learned underlying data from the state of the art and makes accurate predictions autonomously. We also looked into the cases where all the models' predictions align with the DMS\_bin\_score except for our model and found no such cases.
\subsubsection{Case study of LZTR1}
\begin{figure}[!h]
    \centering
    \includegraphics[width=1\linewidth]{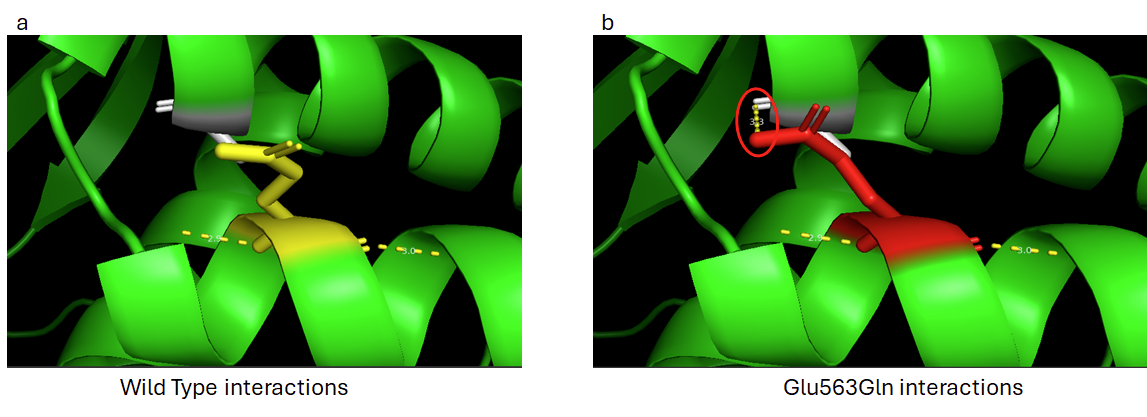}
    \caption{
    \textbf{Visualization of the E563Q mutation in the Leucine Zipper-like Transcriptional Regulator 1 (LZTR1) protein:} 
    \textbf{(a)}: Visualization of the Wild Type residue (Glu563) and its interactions. 
    \textbf{(b)}: Visualization of the mutated residue (Gln563) and its interactions. The Glu563Gln mutation causes a new interaction with an alpha helix (circled in red).}
    \label{fig5}
\end{figure}

For the first case study, we focused on the E563Q mutation in the Leucine Zipper-like Transcriptional Regulator 1 (LZTR1). The Leucine Zipper-like Transcriptional Regulator 1 operates within the Golgi apparatus. It acts as a negative regulator of RAS-MAPK signaling by controlling Ras levels and decreasing Ras association with membranes. LZTR1 is also hypothesized to interact with the CUL3 ubiquitin ligase complex, which facilitates the degradation of redundant proteins. Functionally, LZTR1 is believed to act as a tumor suppressor.

Our model classified the E563Q mutation as pathogenic, which aligns with the experimental annotation from the DMS\_bin\_score. Interestingly, other predictive models, such as AlphaMissense, ESM1b, and GPN-MSA, classified the mutation as benign. Previously, ClinVar had classified this mutation as "likely pathogenic," but a recent update reclassified it as a variant with Conflicting Interpretations of Pathogenicity. In contrast, dbSNP still categorizes this mutation as "likely pathogenic."

To analyze the structural implications of this mutation, we retrieved the protein structure from \href{https://alphafold.ebi.ac.uk/entry/A0A384NL67}{AlphaFold's website} and visualized both the wild-type and mutated sequences using PyMOL\cite{yuan_using_2017}.

As shown in Fig.6, the mutation may lead to the formation of H-bond between N atom of Gln563 and the neighboring alpha helix. In contrast, the wild-type Glu563 cannot form H-bond due to the negatively charged carboxylate of Glu563. This additional interaction between the two alpha helices in the Glu563Gln mutant may reduce the flexibility of the protein’s tertiary structure, potentially altering its function.

Further support for the pathogenicity of the E563Q mutation comes from a study by Johnston et al. \cite{johnston_autosomal_2018} on LZTR1 variants and their role in Noonan syndrome. The study examines a family with the E563Q mutation, where both parents, heterozygous for the mutation, showed no significant phenotypes. However, their two homozygous children displayed severe manifestations of Noonan syndrome. The first child was diagnosed with biventricular hypertrophic cardiomyopathy (HCM) at birth and had distinct facial features, mild short stature, and pectus excavatum. He developed acute lymphoblastic leukemia at age 3 and is now in remission. His younger brother had an atrioventricular septal defect (AVSD), severe biventricular HCM, and a sacral meningomyelocele, and died on day 4 from an inoperable cardiac defect.

These findings, coupled with our structural analysis, suggest that the E563Q mutation in LZTR1 is likely pathogenic, as our model predicted, contrary to the existing state-of-the-art models.

\subsubsection{Case study of KAT6A}

For the second case study, we investigated the E221K mutation in the KAT6A protein.
KAT6A is a histone acetyltransferase responsible for acetylating lysine residues in H3 and H4 histones. As part of the MOZ/MORF complex, KAT6A exhibits histone H3 acetyltransferase activity. It also serves as a transcriptional coactivator for RUNX1 and RUNX2, and acetylates p53/TP53, controlling its transcriptional activity via association with PML.

Our model classified the E221K mutation as benign, which aligns with the DMS\_bin\_score. Clinvar currently classifies the variant as VUS while the state-of-the-art computational models predict it as pathogenic except for Alphamissense which labels it as ambiguous.

The protein does not have a consensus structure, and the Alphafold predictions for the protein show low overall confidence. However, the region surrounding the E221K mutation lies within a high-confidence zone and is located in a coil structure. PyMOL visualization of both WT and mutated residues shows no interaction with neighboring residues (Fig.7), suggesting minimal impact on the protein’s overall structure.

The E221 residue lies within two functional domains. The first one is a Zinc Finger (ZF) (residues in 206-265). The \href{https://prosite.expasy.org/rule/PRU00146}{PRU00146} ZF has no defined function. Zinc fingers typically involve cysteine or histidine residues, while this mutation involves glutamic acid to lysine, which suggests a low likelihood of impacting ZF function. The second domain is an Interaction region with PML (promyelocytic leukemia) containing residues 144-664 \cite{rokudai_moz_2013}. However, structural visualization reveals that the E221K mutation is buried within the protein core and not exposed, making it less likely to participate in significant interactions with adjacent residues \cite{LEE1971379}.

ClinVar initially classified this variant as likely benign but later reclassified it as VUS. According to the authors' submission on ClinVarMiner, this variant has not been reported in individuals affected with KAT6A-related conditions, and the advanced modeling of protein structure and biophysical properties such as structural, functional, and spatial information, amino acid conservation, physicochemical variation, residue mobility, and thermodynamic stability indicate that this missense variant is not expected to disrupt KAT6A protein function. This statement reinforces the hypothesis that such mutation may have a lower impact on the protein function as loops are disordered and may contribute less to the protein function, especially since the mutation's position in 221 does not belong to a functional region.

\begin{figure}[!h]
    \centering
    \includegraphics[width=1\linewidth]{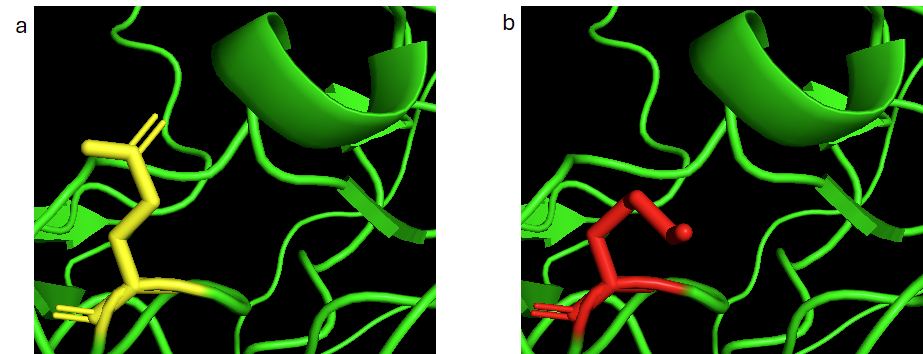}
    \caption{\textbf{Visualization of the E221K mutation in the KAT6A protein.}
    \textbf{(a)}: The wild-type residue (E221) is depicted in yellow, showing no interactions with neighboring structures.
    \textbf{(b)}: The mutated residue (K221) is shown in red, located in a coil region with no interaction with adjacent structures.}
    \label{fig6}
\end{figure}
\section{Conclusion}
This study highlights the potential of combining advanced machine learning models in classifying genetic variants. Our integrated approach consistently outperformed the latest methods in determining the pathogenicity of these variants. The comprehensive evaluation clearly demonstrated the proficiency of the Multi-input Neural Network model in handling both straightforward cases and Variants of Uncertain Significance (VUS).

Our analysis underscored the importance of feature selection. Combining DNA and protein data using potential scores from GPN-MSA and ESM1b established a solid performance baseline. Adding AlphaMissense's scores significantly boosted the predictive power of all models. This validates the advantage of integrating structural insights from protein data with sequence-based predictions. Using both observed and potential scores together led to the best overall results. The hypothesis that emphasizing observed scores would enhance the model's focus on clinically validated mutations proved to be successful. By adding weight to these observed scores, the model better captured real-world genetic variations, leading to improved accuracy and robustness in classification.

Our integrated model consistently showed strong performance across various testing scenarios, demonstrating its capability to effectively interpret complex datasets. ESM1b and AlphaMissense consistently outperformed GPN-MSA. This can be due to the fact ESM1b and AlphaMissense are protein-based models whereas GPN-MSA is trained on DNA data, likely because protein data provides critical structural and functional context necessary for accurate variant classification. The improvements seen when excluding variants classified as ambiguous by AlphaMissense, and the corresponding drop in accuracy for those variants, further emphasize the challenges that uncertain classifications pose to predictive tasks.

In the comparison with ClinVar's annotations, the extended evaluation highlighted the efficiency of our integrated model in distinguishing variant pathogenicity across both the ProteinGym and ClinVar datasets, which are meticulously curated for experimental validity. The model's robust performance, particularly in handling VUS, makes it potentially useful for clinical and research applications where accurate interpretation of ambiguous genetic data is crucial.

Case studies further showcased the practical value of our model's predictions. Investigating protein structures and reviewing related literature supported the accuracy of our model and the DMS\_bin\_score annotations, underlining the model's real-world applicability. This also indicates that our model was effectively trained, capturing underlying information from state-of-the-art models rather than merely replicating their predictions.

 our model is a step forward for characterizing variants of unknown significance and paves the way for identifying new therapeutic targets (or better characterization) or improving models that use NGS data \cite{fossi2024swiftdossiertailoredautomaticdossier, raieli2024escapingforestsparseinterpretable}. Looking ahead, it is essential to extend our validations by testing the integrated model on larger and more diverse datasets. Incorporating additional relevant components, such as transcriptomics scores from SpliceAI\cite{jaganathan_predicting_2019}, could further enhance the model's performance. By integrating these scores, we can add another dimension to our model, combining genomic, proteomic, structural, and transcriptomic data, leading to even more accurate and reliable predictions.

\section{Acknowledgements}
We would like to thank the members of the IT team, Jonathan Schmiedt and Thomas Wursten for their technical assistance during this project.
We would also like to thank Maria Eugenia Riveiro for bringing her invaluable expertise and thorough insights.

\printbibliography 







\end{document}